\input{aipcheck}

\documentclass[
    ,final            
  ]
  {aipproc}

\layoutstyle{8x11single}

\def\sun{\hbox{$\odot$}} 
\def\degr{\hbox{$^\circ$}} 

\begin{document}

\title{Massive stars in colliding wind systems:\\
 the GLAST perspective}


\author{Anita Reimer \& Olaf Reimer}{
  address={Stanford University, HEPL \& KIPAC, Stanford, CA 94305, USA}
}

\begin{abstract}
Colliding winds of massive stars in binary systems are considered as candidate sites of high-energy non-thermal photon emission. They are already among the suggested counterparts for a few individual unidentified EGRET sources, but may constitute a detectable source population for the GLAST observatory.

The present work investigates such population study of massive colliding wind systems at high-energy gamma-rays. Based on the recent detailed model (Reimer et al. 2006) for non-thermal photon production in prime candidate systems, we unveil the expected characteristics of this source class in the observables accessible at LAT energies. Combining the broadband emission model with the presently cataloged distribution of such systems and their individual parameters allows us to conclude on the expected maximum number of LAT-detections among massive stars in colliding wind binary systems.
\end{abstract}

\maketitle


\section{Introduction}

Massive stars are hot, high-mass and luminous ($\sim 10^{5\dots 6}L_{\sun}$)
stellar systems that possess supersonic winds ($V_{{\rm W},\infty}\sim$ several 1000~km/s) and accompagnied with a large mass loss rate ($\dot M\sim 10^{-6\ldots -4}M_{\sun}$/yr). From more than a dozen galactic Wolf-Rayet (WR) binary systems non-thermal radio emission has been observed \cite[e.g.,][]{Abbott1986}, and this is interpreted as synchrotron radiation from a relativistic electron distribution. As a consequence inverse Compton (IC) scattering in the dense photospheric radiation fields of the stars as the presumably dominant leptonic emission process \cite{Eichler1993,Muecke2002} is guaranteed to produce high-energy photons. Observational indications for massive stars systems being high-energy sources comes from population studies of the still unidentified EGRET sources: correlations with various massive star populations have been found \cite[e.g.,][]{Romero1999}, although their unambiguous identification must await instruments with significantly improved source localization capabilities (e.g. GLAST). 
The site of particle acceleration may be the instability driven shocks inside the winds \cite{White1985}, shocked wind collision regions of multiple systems \cite{Eichler1993,Bykov1992} or the termination shock \cite{Voelk1982}. The following work concerns colliding winds (CWs) of WR-binary systems. Its main characteristics \cite{Eichler1993} has been meanwhile impressively confirmed by radio as well as X-ray observations \citep[see e.g.,][for the nearby WR~147 system]{Dougherty2002,Pittard2002}:
The collision of supersonic winds is creating a region of hot shocked gas that is separated by a contact discontinuity, and a forward and reverse shock follows. Its stagnation point is defined by the balance of the wind momenta (assuming homogeneous winds). The observation of synchrotron radiation indicates that the winds are permeated by magnetic fields supposedly originating from the massive star's surface. Estimates for surface magnetic
field strengths range from below $B_s=100$~G (e.g. \cite{Mathys1999}) up to $\sim 10^4$G 
\cite{Ignace1998}. In the following we fix this value to a reasonable 100~G, unless 
stated otherwise, and use the magnetic rotator theory (e.g. \cite{Weber1967}) to estimate
the field strength at the CWR. Typically $>$mG or higher 
field strengths are expected, assumed to
be constant throughout the emission region. 
Note that because the stellar target photons for IC scattering arrive at the collision region from a 
preferred direction, the full angular dependence of the cross section has to be taken into account 
(e.g. \cite{Reynolds1982}).
This leads to anisotropy effects like the emitted flux and cutoff energy 
being dependent on the sight line into the wind \cite[see e.g.,][]{Reimer2006}. In binary systems the
maximum flux level will be reached at phases where the WR- is behind the OB-star along the line-of-sight. Photon absorption due to $\gamma\gamma$ pair production may affect the spectrum at energies above $\sim 70$~GeV in CW systems.

The most up-to-date catalogs of galactic WR- \cite{WRcat,WRcatextension} and OB-stars \cite{Maiz2004}
contains more than
220 and 370 sources, respectively, with a detected binary fraction of $\sim 40-50\%$ \cite{WRcat}. More than 30\% are located in clusters and associations. 
Its spatial distribution reflects the spiral arm structure of our Galaxy (\cite{WRcat} and references therein).
 The total number of
WR-(systems) in the Galaxy has been estimated to $\sim 8000$ \cite{Prantzos1986},
total OB-star number counts in our Galaxy may reach values as high as 60000 \cite{Reed2000}.
This is at 
least an order of magnitude larger than the number of all presumably galactic, but
still unidentified $\gamma$-ray sources detected to date.
How many of those massive WR-binary systems will GLAST's LAT be detecting {\bf at most}? By combining a 
realistic physical source model with the statistics offered by the star catalogs we aim to 
answer this question in the following. The result will have implications for the chosen approach to 
probe the existence of WR-binaries in the the GLAST data. Whereas spatial coincidences are expected given 
the number of cataloged Wolf-Rayet stars and the anticipated number of $\gamma$-ray source detections  
in the Galactic Plane above GLAST sensitivity, this study provides a prior for the size of a plausible gamma-ray bright 
WR-binary population as well as the individuals to be preferably investigated among the population.

\section{Sample selection and parameters}

The construction of the sample for our study starts with the galactic WR-catalog \cite{WRcat,WRcatextension}, selecting all 88 binary systems.
For more than a dozen of those sources the nominal shock location falls below the star's photosphere
if wind-momentum balance determines the shock location, making the CW scheme an unsuitable setting
for those. These are excluded in the following. Because of the distance squared dilution factor, close-by sources have a higher probability to
be detected. Indeed, our statistical study shows that sources beyond $\sim 3$~kpc have a negligible chance
to become a LAT source. Hence a distance cutoff in our sample was chosen at 4~kpc, which excludes a further 42 systems. 
Finally, a
$\gamma$-ray flux estimate on the basis of model simulations requires at least the knowledge of the shock location, estimated from its period or stellar separation and wind momenta of the participating stars, to provide reasonable estimates for the photon and 
field density at the shock. More than 10 sources have no period assigned, and hence are excluded from 
this study.
We end up considering 21 galactic WR-binaries for potential
LAT detectability.

For estimating the maximum possible flux in LAT's energy range we consider 
in the following the
IC component only as the most dominant $\gamma$-ray production process, and use the Reimer et al. (2006)
WR-binary model. The most important system parameters that impact the flux estimate significantly are
the photospheric luminosity and temperature of the stars, their masses and momentum loss rates, their stellar separation and system distance. Only the distribution of the two latter ones shows an appreciable 
spread whereas all others have rather narrow ones, owing to the definition of a massive star.
System parameters have been derived from \cite{WRcat,Markova2005,Nugis2000,Schaerer1992}.
Because we're seeking the maximum number of likely LAT-sources among the WR-binary population, we allow
the theoretically highest possible acceleration rate ($\dot E\propto Ec/R_L(V_W/c)^2$, $R_L=$ Larmor radius) and evaluate the $\gamma$-ray output at a phase of maximum IC-flux. The system eccentricity $e$ is known for only half a dozen sources in the considered sample with an average $e\approx 0.2$ and the highest being $e\approx 0.9$. In the following we assume $e=0$, and evaluate then the resulting $\gamma$-ray flux uncertainty using extreme eccentricities. The same applies
to the surface field, fixed at $B_s=100$~G now. An inclination angle of $90\degr$ is used for unknown system inclinations. Particle number and energy conservation arguments
apply for the electron energy injection, as outlined in \cite{Reimer2006}. An particle injection 
spectrum $\propto E^{-2}$ is used throughout this work. With these parameters we calculate the maximum flux expected at $>100$~MeV for all sample sources
using the model of \cite{Reimer2006}.

\section{Results}

We consider a source to be detectable by the LAT if the maximum photon energy reaches at least the LAT 
energy range, and if the photon flux $>100$MeV is above LAT's sensitivity. For the latter
we use an estimated factor 5 higher sensitivity level as published for LAT's high-latitude
sensitivity\footnote{\url{http://www-glast.slac.stanford.edu/software/IS/glast_lat_performance.htm}}, $2\times 10^{-8}$ cm$^{-2}$ s$^{-1}$,
to account the fact that for the massive star systems being located preferentially in the Galactic Plane.
Among the considered sample we find 6-7 WR-binaries to be detectable by the LAT in an one-year
exposure, of which three are positional coincident with EGRET unidentified sources. All but one (for WR~70 only an upper limit in the radio band exists) turn out to be non-thermal 
radio emitters, and tend to be preferentially very-long-period binaries (see Fig.~1). 
The ratio of the wind momenta together with the stars' separation determines the distance of the
shock $x_{\rm OB}$ above the photosphere. The resulting distribution of the
the shock-star separation (see Fig.~1) highlights the preference of LAT-detectible sources to possess
larger values of $x_{\rm OB}$. This is plausible in the light of the severe IC-losses causing a low energy
cutoff in the electron spectrum if the shock is located too deep in the dense photospheric
radiation field, and thus inhibiting GeV photon production in shorter-period binaries.
Finally, we find that only close-by systems ($<1$kpc) have a high chance to be detected by the LAT,
whereas WR-binaries located beyond $\sim 1$~kpc but up to $\sim 3$~kpc have a significantly reduced detection probability (see Fig.~1). 

To estimate the robustness of our result in view of unavoidable uncertainties in the parameter
values, we vary each parameter (distance $d_L$, stellar separation $D_{\rm WR-OB}$, surface magnetic field $B_*$, 
mass loss rates $\dot M_{\rm WR}$, wind velocities $V_{\infty,\rm WR,OB}$, stellar temperature
$T_{\rm eff}$ and luminosity $L_{\rm bol}$) and deduce the resulting number of LAT-detectable
sources after each change. Specifically we alter period, $d_L$, $B_*$, 
$\dot M_{\rm WR,OB}$, $V_{\infty,\rm WR,OB}$, $T_{\rm eff}$ and $L_{\rm bol}$  by a factor
2, 2, 10, 10, 2, 0.2 and 2 respectively. A variation of the system eccentricity to the extreme possible values adds a further factor $\pm^{1.9}_{0.1}$ uncertainty to the stellar separation.
As a result we find 6-7 systems in our sample are LAT-detectable with a total uncertainty of $\pm^2_5$, leading to the conclusion that the expected maximum number of detectable WR-binaries in a one-year GLAST-LAT exposure may be as high as eight, but
could be as low as only one source from this population. 

Identification procedures may substancially benefit from physically motivated estimates of the expected source density of the respective 
population, like suggested here for the case of WR-binaries. Further parameter studies will include their spatial distribution, flux variability and orbital periodicity, spectral features, and signatures obtained at other wavelengths
(see \cite{Reimer2007} for a discussion applied to massive star systems).

\begin{figure}
  \includegraphics[height=.2\textheight]{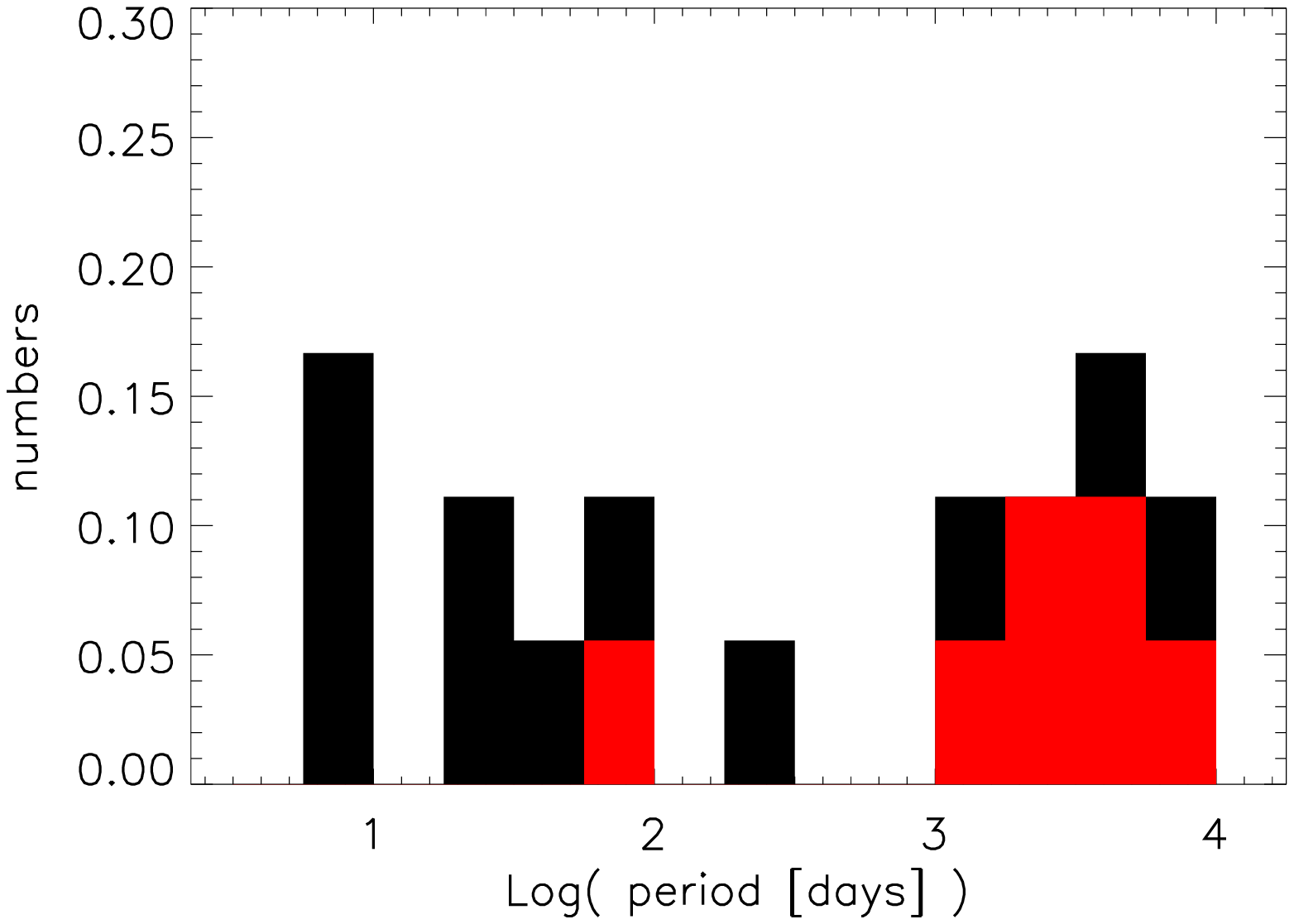}
  \includegraphics[height=.2\textheight]{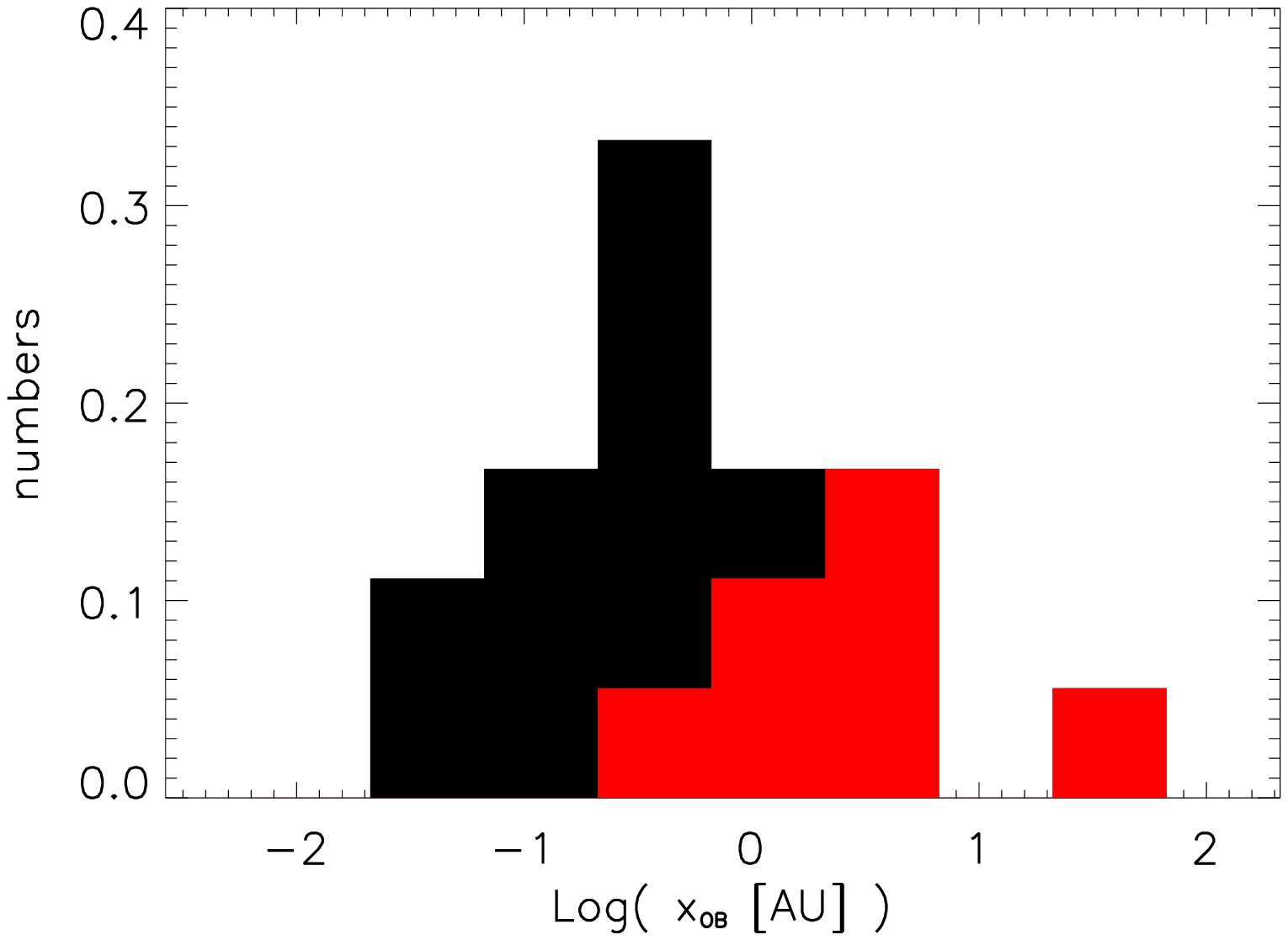}
  \includegraphics[height=.2\textheight]{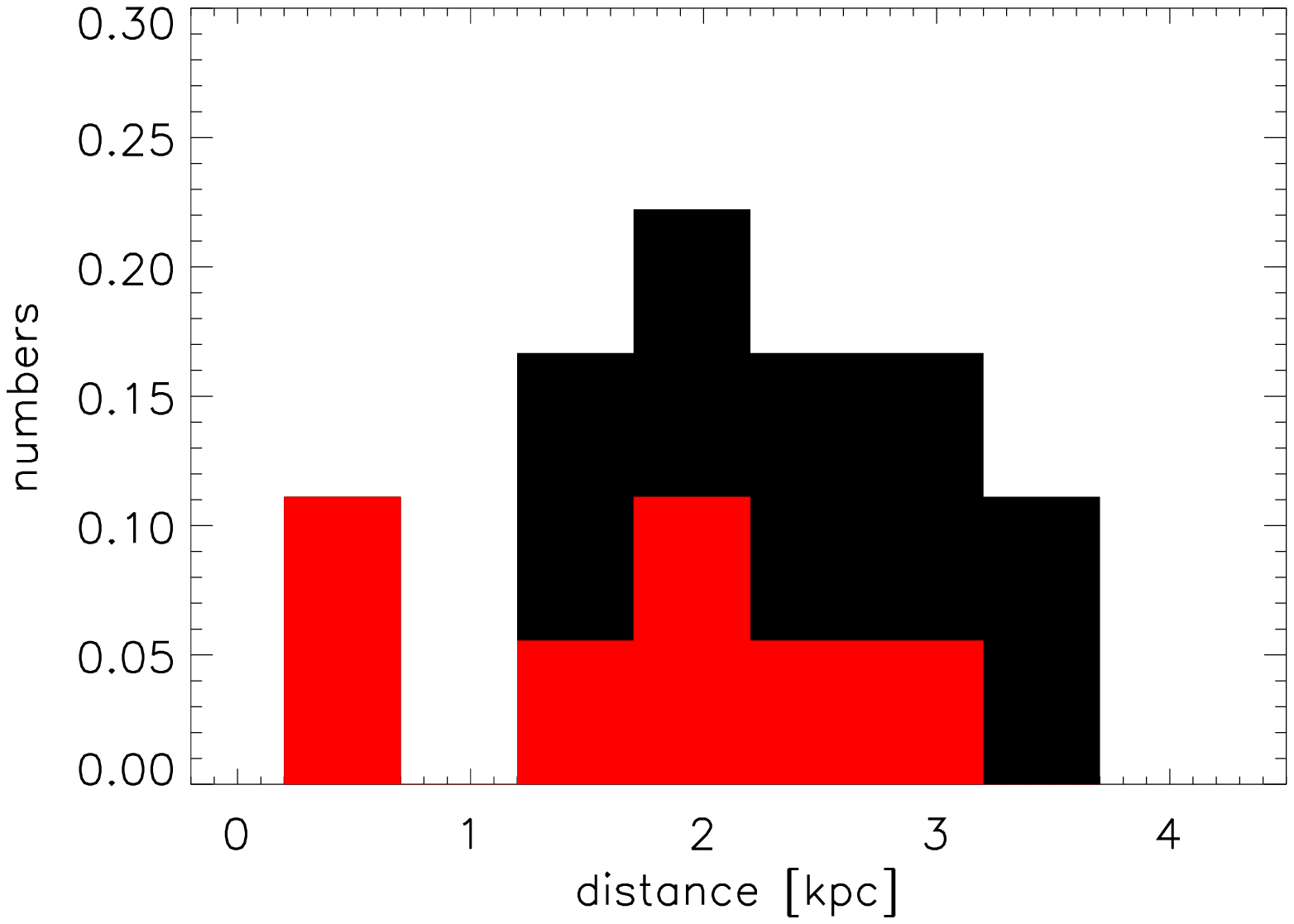}
  \caption{Normalized source number distribution of period (left), shock-star distance $x_{\rm OB}$ (middle) and source distance (right) of LAT-detectable WR-binaries (red/grey) from the source sample in comparison to all sample sources (red/grey and black). 
\vspace*{-.3cm}}
\end{figure}





\bibliographystyle{aipproc}   


\IfFileExists{\jobname.bbl}{}
 {\typeout{}
  \typeout{******************************************}
  \typeout{** Please run "bibtex \jobname" to optain}
  \typeout{** the bibliography and then re-run LaTeX}
  \typeout{** twice to fix the references!}
  \typeout{******************************************}
  \typeout{}
 }




\end{document}